\documentclass[
prd 
,showpacs, amssymb, superscriptaddress, aps, 
]{revtex4}
\usepackage{graphicx}
\input{epsf}

\usepackage{amsmath,amssymb}
\usepackage{bm}
\usepackage{times}
\usepackage{ulem}

\newcommand{\dalm}{\kern1pt\vbox{\hrule height 0.9pt\hbox{\vrule width 0.9pt
\hskip 2.5pt\vbox{\vskip 5.5pt}\hskip 3pt\vrule width 0.3pt}\hrule height 0.3pt}
\kern1pt}

\newcommand{\gsim}{\, \raisebox{-0.8ex}{$\stackrel{\textstyle >}{\sim}$ }}
\newcommand{\lsim}{\, \, \raisebox{-0.8ex}{$\stackrel{\textstyle <}{\sim}$ }}

\begin{document}



\title{Dimension dependence of numerical simulations on gravitational waves from protoneutron stars}

\author{Hajime Sotani}
\email{sotani@yukawa.kyoto-u.ac.jp}
\affiliation{Astrophysical Big Bang Laboratory, RIKEN, Saitama 351-0198, Japan}
\affiliation{Interdisciplinary Theoretical \& Mathematical Science Program (iTHEMS), RIKEN, Saitama 351-0198, Japan}

\author{Tomoya Takiwaki}
\affiliation{Division of Science, National Astronomical Observatory of Japan, 2-21-1 Osawa, Mitaka, Tokyo 181-8588, Japan}
\affiliation{Center for Computational Astrophysics, National Astronomical Observatory of Japan, 2-21-1 Osawa, Mitaka, Tokyo 181-8588, Japan}

\date{\today}

\begin{abstract}
We examine the eigenfrequencies of gravitational waves from the protoneutron stars (PNSs) provided via the core-collapse supernovae, focusing on how the frequencies depend on the dimension of numerical simulations especially for the early phase after core bounce. As expected, we find that the time evolution of gravitational wave frequencies depends strongly on the dimension of numerical simulation as well as the equation of state for high density matter. Even so, we find that the fundamental frequency as a function of PNS average density and the ratio of the specific eigenfrequencies to the fundamental frequency as a function of PNS properties are independent of the dimension of the numerical simulations, where the dependence of the equation of state is also very weak in this early postbounce phase. Thus, one can safely discuss the gravitational wave frequencies of the PNSs as a function of the PNS average density or compactness even with the frequencies obtained from the one dimensional simulations. We also provide phenomenological relation between the compactness and average density as well as the relation among the $f$-, $p$-, and $g$- mode frequencies.
\end{abstract}

%
\maketitle


\section{Introduction}
\label{sec:I}

Direct detection of gravitational waves from the compact binary mergers opened a new window for multimessenger era, where gravitational waves can become fruitful for obtaining astronomical information together with electromagnetic waves and neutrinos. In fact, at the event of GW170817, the electromagnetic counterparts have also been detected, following the detection of gravitational waves \cite{GW6,EM}. In addition to the Laser Interferometer  Gravitational-Wave Observatory (LIGO) and Virgo, the Japanese gravitational wave detector, KAGRA \cite{aso13}, will operate soon, while the discussion for the third-generation detectors, such as the Einstein Telescope \cite{punturo} and Cosmic Explorer \cite{CE}, have already been started. Via such most-advanced detectors, it is expected that many gravitational wave events will be observed in near future. At that time, not only the gravitational waves from compact binary mergers but also that from supernova explosions might be detected. Since the supernova explosion is much less energetic source for gravitational waves compared to the compact binary mergers, the detection may be possible for only the event very close to us, i.e., up to $\sim 10$ kpc (e.g., \cite{Hayama15,Andresen2018}). Even so, if it would come true, one can expect to simultaneously detect not only the gravitational waves together with the electromagnetic waves but also the neutrinos.

So far, the gravitational waves from supernova explosions have been discussed extensively with the numerical simulations of core-collapse supernova (e.g., \cite{Murphy09,MJM2013,Ott13,CDAF2013,Yakunin15,KKT2016,Andresen16,OC2018,RMBVN19} for non-rotating models and \cite{Dimmelmeier2008,Takiwaki2010,Reisswig2010,Kotake2011,Ott2012,Kuroda2013,Yokozawa2014,Takiwaki2017,Richers2017,Andresen2018} for rotating models). In these attempts, the existence of gravitational wave signals has been obviously shown, where the frequency of the strongest gravitational wave signal increases with time from a few hundred hertz up to kilohertz. This signal is also characterized by the protoneutron star (PNS) properties \cite{MJM2013,CDAF2013}, i.e., the frequency is identified as a surface $g$-mode oscillation of PNSs calculated via the Brunt-V\"{a}is\"{a}l\"{a} frequency at the PNS surface.

The numerical simulations are definitely essential for understanding some aspects of the mechanism of supernova explosions, but they may be sometimes unsuitable for parameter search and/or extraction of physics behind the numerical data due to high computational costs in numerical simulations. In order to complement this difficulty, the perturbative approach is another powerful technique. Since the oscillation frequency of astronomical objects strongly depends on their interior properties, as an inverse problem one would extract the interior information via detection of such a frequency. This technique is known as asteroseismology, which is very similar to seismology in Earth and helioseismology in Sun. In practice, asteroseismology for cold neutron stars has been frequently discussed up to now. For example, it is reported that the neutron star crust properties are constrained by identifying the quasi-periodic oscillations observed in the giant flare activities as the crustal torsional oscillations \cite{SW2009,GNHL2011,SNIO2012,SIO2016}. In addition, via the observations of gravitational waves, one may obtain the mass, radius, equation of state (EOS), and rotational properties of compact objects (e.g., \cite{AK1996,AK1998,STM2001,SH2003,SYMT2011,PA2012,DGKK2013}).

Compared to the cold neutron stars, the studies about asteroseismology in PNSs (or hot neutron stars) are still poor. This may be because that, unlike cold neutron stars, one needs to know the profiles of electron fraction and entropy as well as that of pressure as a function of density, which can be obtained only after numerical simulation of supernova explosions. Even so, owing to the development of numerical simulations, it is becoming that the discussions of asteroseismology in PNSs are done gradually  \cite{MRBV2018,SKTK2019,TCPF2018,TCPOF2019a,TCPOF2019b,SS2019,FMP2003,Burgio2011,FKAO2015,ST2016,Camelio17,SKTK2017,VBR2019,WS2019}. In these studies one needs to prepare the background PNS models for perturbation analysis, where two different approaches have been considered. One is that the PNS surface is determined with a specific density and one considers the specific oscillations inside such PNS models (e.g., \cite{MRBV2018,SKTK2019,SS2019,FMP2003,Burgio2011,FKAO2015,ST2016,Camelio17,SKTK2017,VBR2019}). In this case, there may be an uncertainty for selection of the surface density, but at least the dependence of $f$-mode frequency on the surface density seems to disappear $\gsim 500$ ms after the core bounce \cite{MRBV2018}. Another approach is that one considers the numerical domain inside the shock radius and calculates the specific oscillations inside such a numerical domain (e.g., \cite{TCPF2018,TCPOF2019a,TCPOF2019b,WS2019}). With this approach, one can avoid the uncertainty for the selection of outer boundary, but turbulent motions especially outside the PNS region are more violent, which may be a problem for preparing the spherically symmetric background model. In addition, the boundary condition imposed at the outer boundary, i.e., at the shock radius, is not the same as that in standard asteroseismology. As a result, one has to reclassify the oscillation modes with a kind of new definition. In this paper, we simply adopt the PNS models, whose surface are determined with a specific density, and will discuss the PNS asteroseismology.

Anyway in the both perturbative approaches, to calculate the specific frequencies of gravitational waves from the system, one has to prepare the background models with using the numerical data obtained by the numerical simulation of core-collapse supernovae. To prepare the spherically symmetric background models, the numerical data obtained via multi-dimensional simulations is averaged in the angular direction(s). So far, the PNS asteroseismology has been discussed with the several numerical data obtained by various numerical simulations with different dimensions and with various progenitor models. However, it has never been systematically discussed how the gravitational wave spectra can depend on the dimension of numerical simulations for preparing the PNS models. So, in order to clarify this point, we will study the PNS asteroseismology in this article with using the PNS models obtained via the one-dimensional (1D), two-dimensional (2D), and three-dimensional (3D) numerical simulations with the same progenitor models. Namely,  we make the numerical simulations with different dimensions based on the same numerical scheme, and then determine the specific frequency on the PNSs obtained with such numerical simulations.

This paper is organized as follows. In Sec. \ref{sec:PNSmodel}, we describe the PNS models and their properties considered in this study. In Sec. \ref{sec:Oscillation}, we report the eigenfrequencies of gravitational waves, especially focusing on the dependence of the dimension of the numerical simulations for preparing the PNS models. We also show the dependence of eigenfrequencies in gravitational waves on the PNS properties. Finally, we summarize in Sec. \ref{sec:Conclusion}. Unless otherwise mentioned, we adopt geometric units in the following, $c=G=1$, where $c$ denotes the speed of light, and the metric signature is $(-,+,+,+)$.

\section{PNS Models}
\label{sec:PNSmodel}

The background PNS models are obtained by neutrino radiation hydrodynamic simulations in {\it 3DnSNe} code (see \cite{takiwaki2016,oconnor2018,kotake2018,nakamura2019,sasaki2019,zaizen2019} for recent applications). We performed 1D, 2D, and 3D simulations, changing the dimensionality. The choice of the resolution of the spherical polar grid is $512$ for 1D run , $512\times128$ for 2D run,  and $512\times64\times128$ for 3D run. The radial grid ($r$) covers from the center to an outer boundary of $5000\mathrm{~km}$. Our spatial grid ($dr$) has a finest mesh spacing of 0.25 km at the center and ${\rm d}r/r$ is better than 2 percent at $r \ge 100\ {\rm km}$. Neutrino transport is solved by isotropic diffusion source approximation \cite{liebendoerfer2009,takiwaki2014}. The employed set of neutrino reaction is same as that in Ref. \cite{kotake2018} and the update on the neutrino nucleon scattering is also  included \cite{horowitz2017}. The evolution of the hydro and neutrino radiation quantities, e.g., the shock radius, neutrino intensities, and average energies, in our code is similar to that shown in Ref. \cite{oconnor2018}.

In order to see the EOS dependence, we adopt two different EOSs in this study. One is DD2 based on the relativistic mean field approach~\cite{DD2}, while another is the EOS (hereafter referred to as TGTF) derived with variational method~\cite{Togashi17} (TF means Thomas-Fermi approximation used in the low density part of EOS). The incompressibility ($K_0$) of symmetric nuclear matter, the symmetry energy ($S_0$), the slope parameter $(L)$, and the maximum mass ($M_{\rm max}$) of the corresponding cold neutron star are respectively $K_0=243$ MeV, $S_0=31.67$ MeV, $L=55.04$ MeV, and $M_{\rm max}=2.42M_\odot$ for DD2, while $K_0=245$ MeV, $S_0=29.1$ MeV, $L=38.7$ MeV, and $M_{\rm max}=2.21M_\odot$ for TGTF. That is, DD2 is stiffer than TGTF. We remark that the both EOSs are not excluded from the observations, i.e., the $2M_\odot$ neutron stars \cite{Demorest2010,Antoniadis2013} and the radius of a $1.4M_\odot$ neutron star constrained from GW170817 \cite{Annala18}. With such two EOSs, we will consider the PNS models obtained by 1D, 2D, and 3D simulations, where the numerical results obtained by 2D and 3D simulations are additionally averaged in the angular direction(s) to prepare the spherical PNS models for linear analysis. Hereafter, we refer to the PNS models by combining the adopted EOS and the dimension of numerical simulation, such as DD2-1D for the PNS model constructed with DD2 and obtained via 1D simulation.

We adopt progenitor of $20M_\odot$ model given in Ref. \cite{WH07}, that is widely used for the core-collapse simulations \cite{bruenn2013,dolence2015,summa2016,suwa2016,pan2016,skinner2016,bruenn2016,oconnor2018b,just2018,glas2019}. The hydrodynamic evolution of the models is similar to the recent calculation with similar setup of neutrino reaction rates \cite{melson2015,bollig2017,vartanyan2018}. We found shock revival time around 300 ms in all 2D and 3D models. In 1D run, the revival is not observed. The dynamics would be discussed in detail in the forthcoming paper.
In this study, we particularly consider the early phase after core bounce, i.e., we consider the numerical results until 320 ms after core bounce, which is restricted by the difficulty for making a long-term simulation in 3D.

\begin{figure*}[tbp]
\begin{center}
\begin{tabular}{ccc}
\includegraphics[scale=0.43]{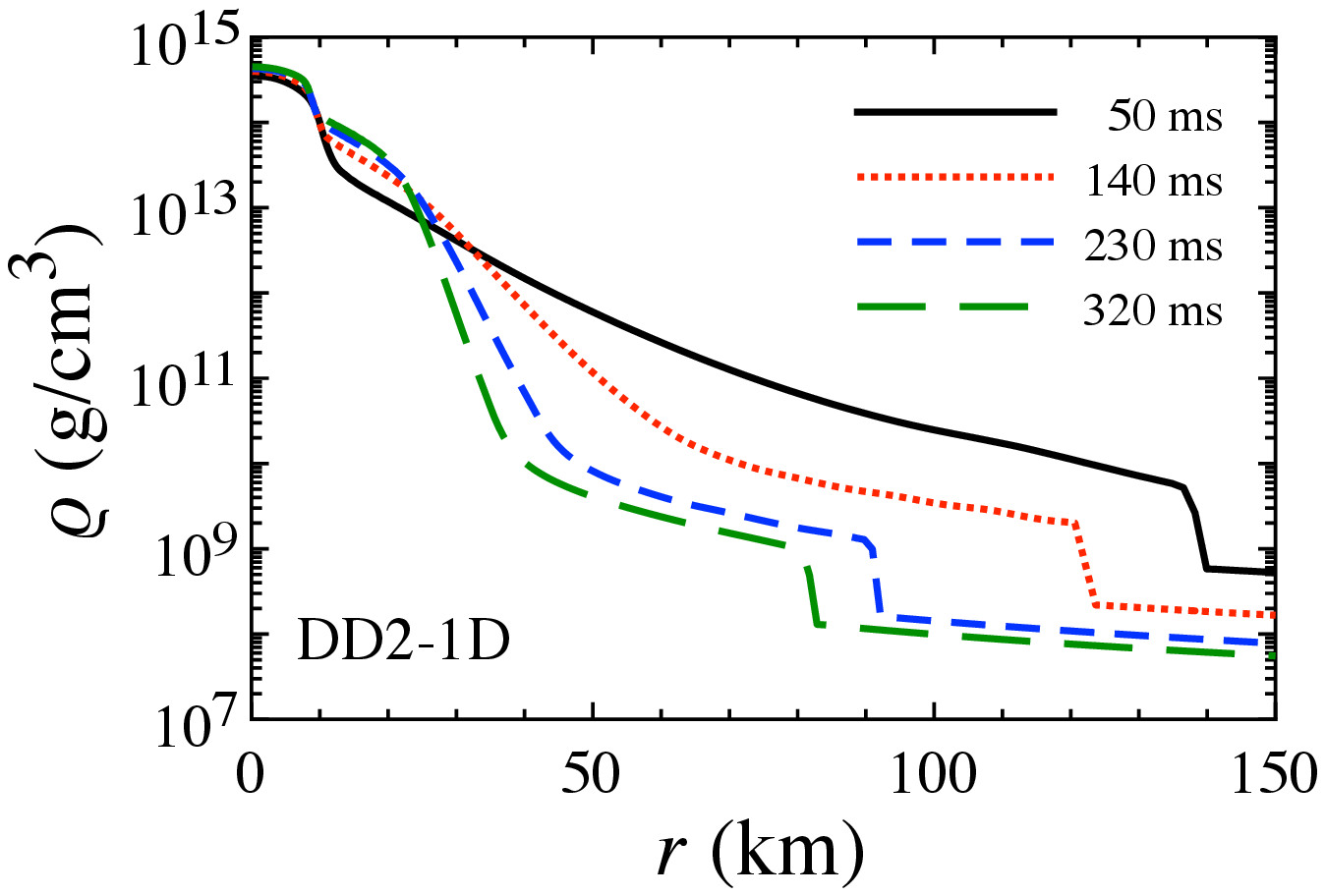} &
\includegraphics[scale=0.43]{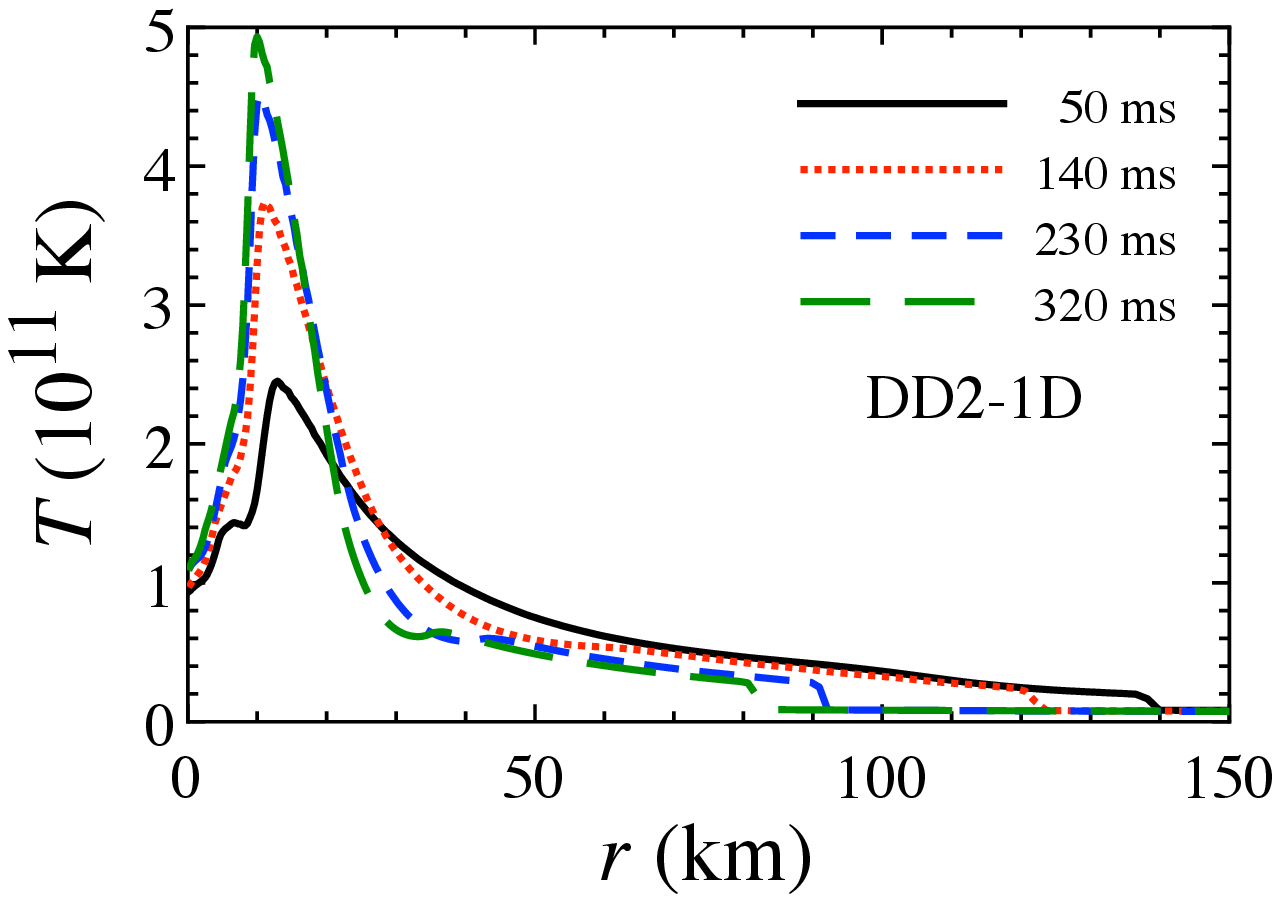} &
\includegraphics[scale=0.43]{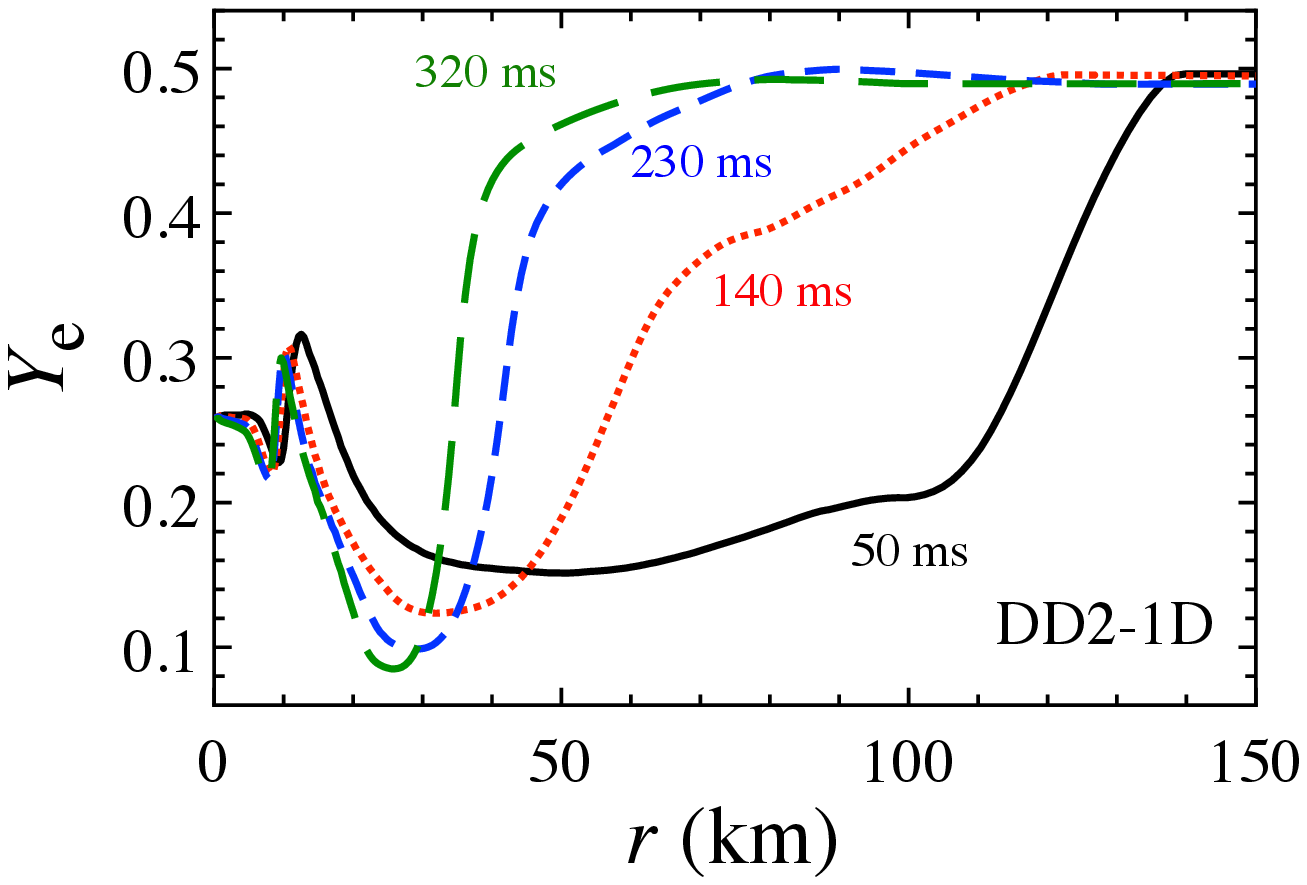} 
\end{tabular}
\end{center}
\caption{
The profiles of rest-mass density (left panel), temperature (middle panel), and electron fraction (right panel) are shown for the PNS model of DD2-1D at $T_{\rm pb}\simeq 50$, 140, 230, and 320 ms. 
}
\label{fig:DD2-1D}
\end{figure*}

Then, the radial profile of the selected time snapshots in the simulations is used as a background model for calculating the eigenfrequency of the oscillation mode.
As an example, in Fig. \ref{fig:DD2-1D} we show the radial profiles of rest-mass density ($\rho$), temperature ($T$), and electron fraction ($Y_{\rm e}$) for the PNS models of DD2-1D for various core bounce time ($T_{\rm pb}$). In this study, we especially adopt $\rho_s=10^{11}$ g/cm$^3$ as the surface density as in Refs. \cite{SKTK2019,SS2019}. From this figure, one can observe that the temperature ($T$) and electron fraction ($Y_{\rm e}$) at the PNS surface for DD2-1D are almost constant during the period we considered, where $T\sim 5\times 10^{10}$ K and $Y_{\rm e}\sim 0.15$, respectively.

\begin{figure*}[tbp]
\begin{center}
\begin{tabular}{ccc}
\includegraphics[scale=0.43]{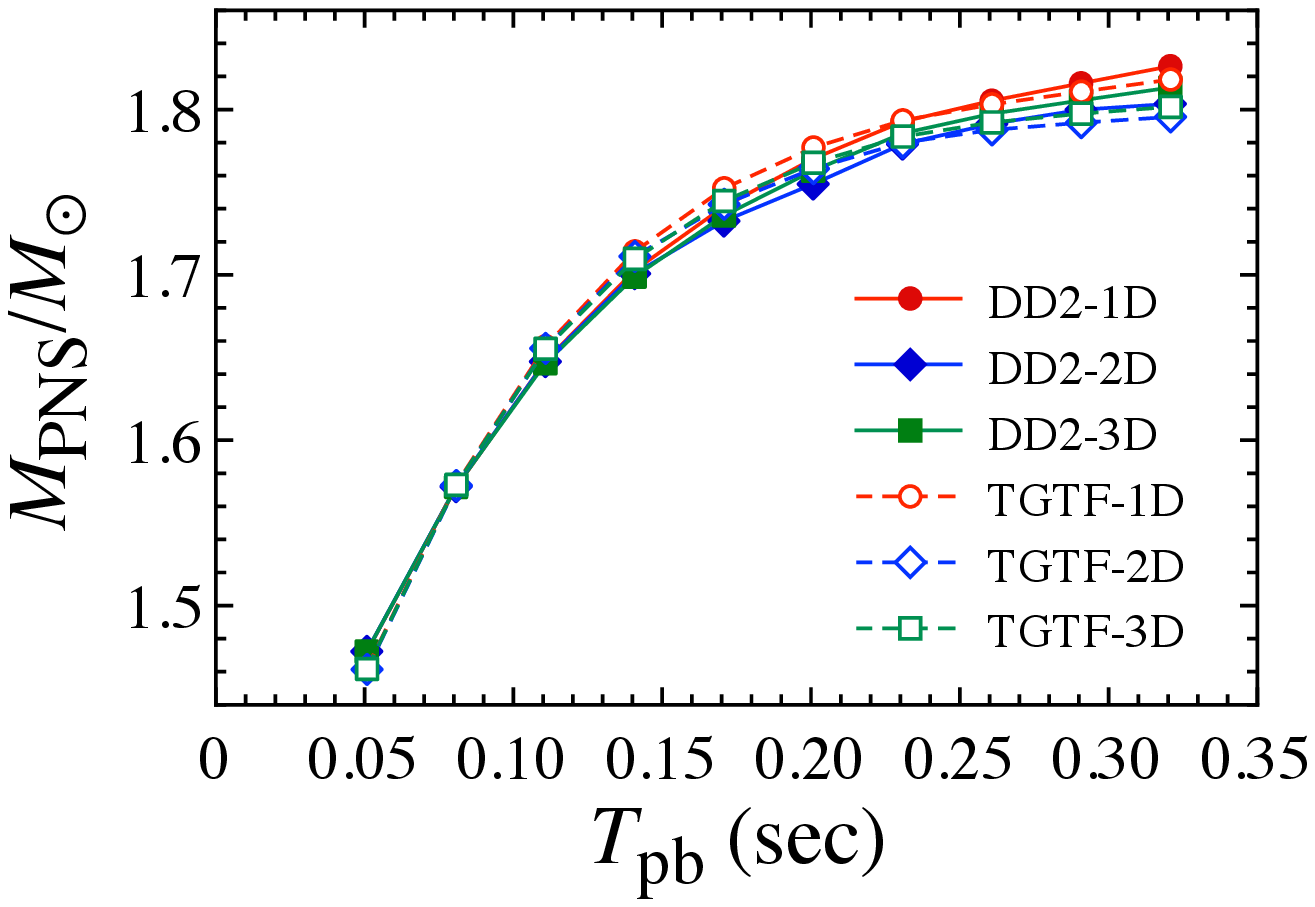} &
\includegraphics[scale=0.43]{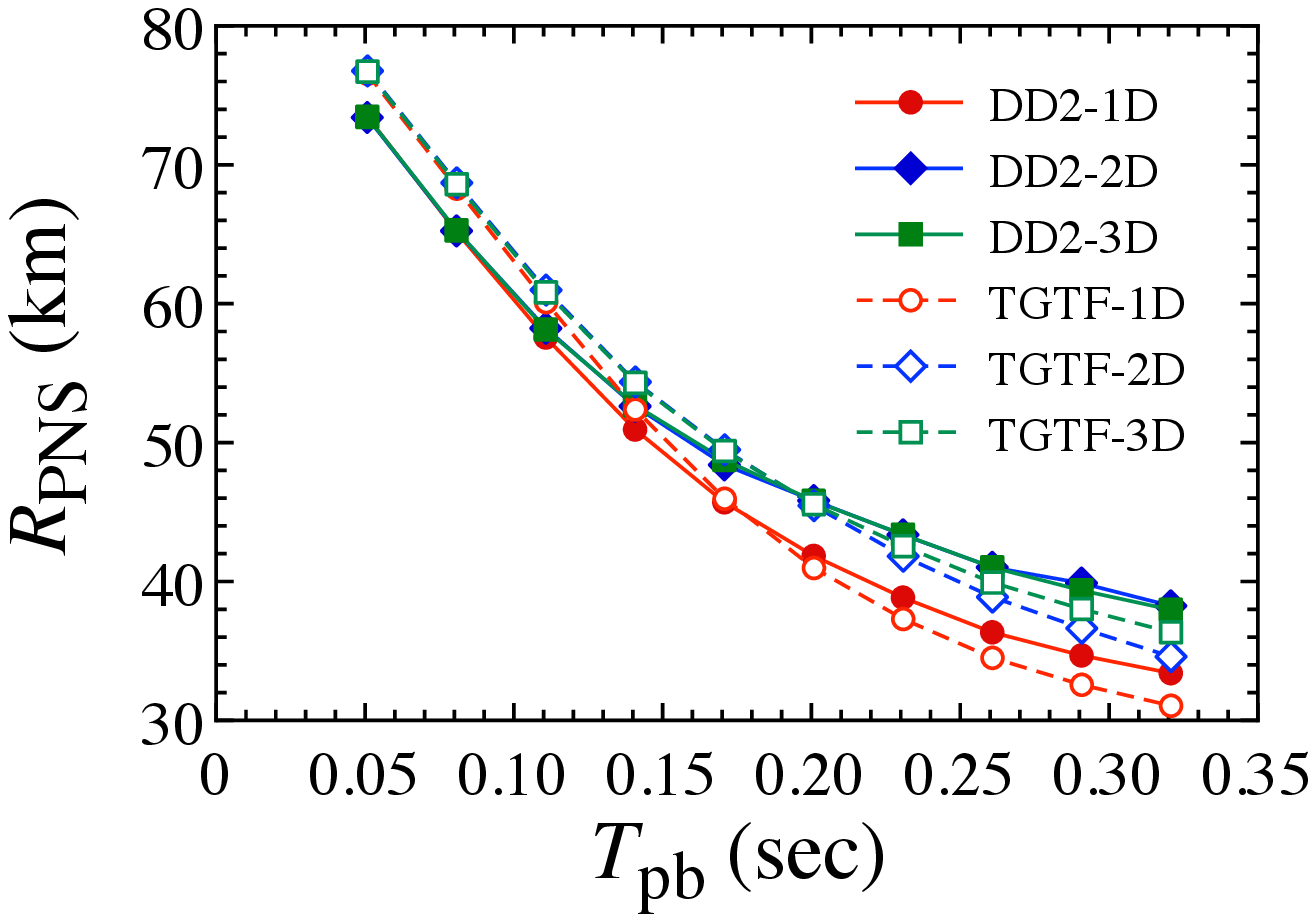} &
\includegraphics[scale=0.43]{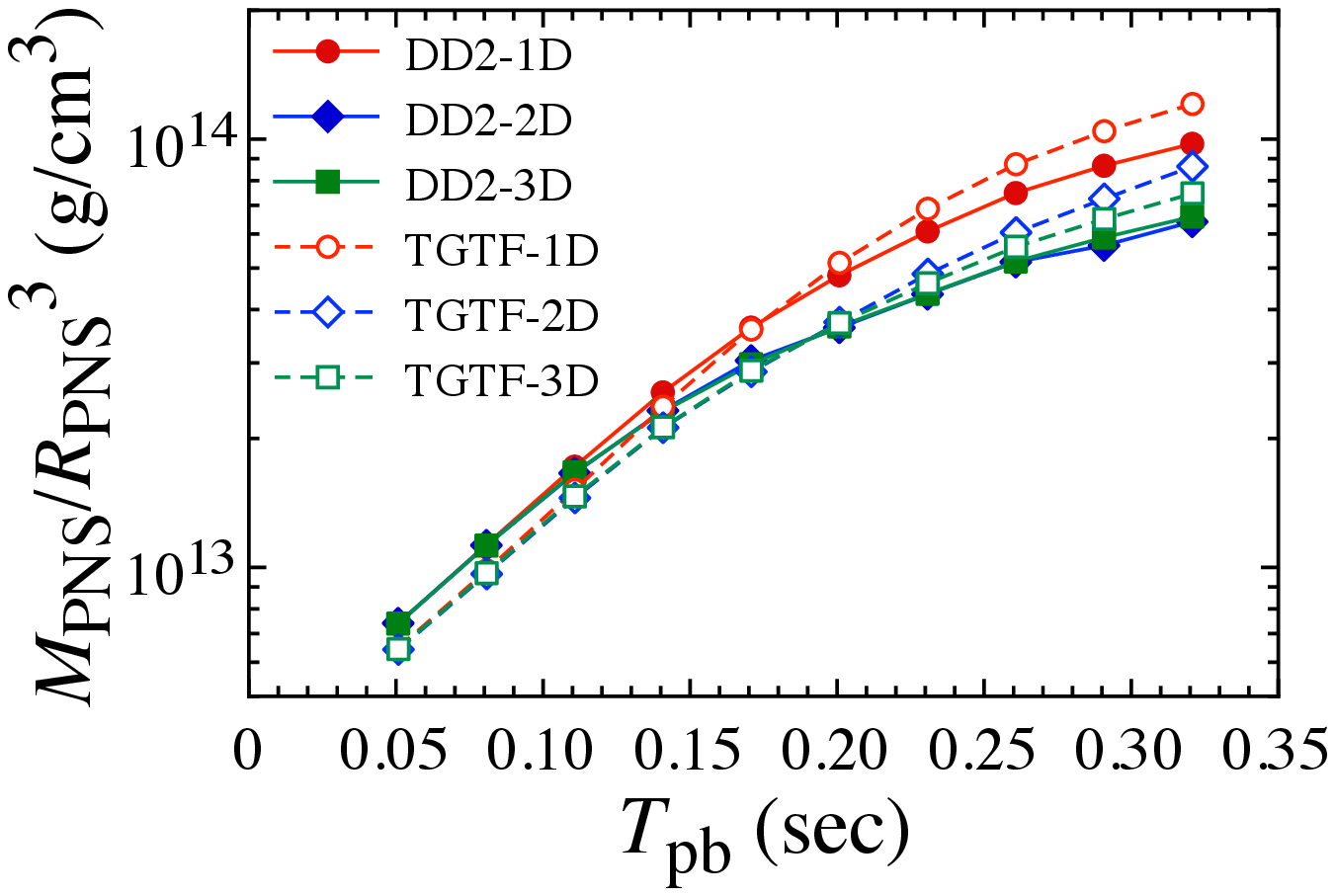} 
\end{tabular}
\end{center}
\caption{
Mass (left panel), radius (middle panel), and average density (right panel) of PNSs obtained from the numerical simulations with different dimensions, are shown as a function of the time after the core bounce, where circles, diamonds, and squares correspond to the results with 1D, 2D, and 3D simulations, respectively. In addition, the filled marks with solid lines denote the PNS models with DD2, while the open marks with dashed lines denote the PNS models with TGTF. The PNS surface is determined with the surface density of $10^{11}$ g/cm$^3$.
}
\label{fig:MRave}
\end{figure*}

The mass ($M_{\rm PNS}$) and radius ($R_{\rm PNS}$) of PNS can be calculated with the fixed surface density. In Fig.~\ref{fig:MRave}, we show the time evaluations of $M_{\rm PNS}$, $R_{\rm PNS}$, and average density ($M_{\rm PNS}/R_{\rm PNS}^3$) of PNSs considered in this study, where the circles, diamonds, and squares correspond to the results with 1D, 2D, and 3D simulations, respectively. In this figure, the filled marks with solid lines denote the PNS models with DD2, while the open marks with dashed lines denote the PNS models with TGTF. In general, the PNS mass increases due to the mass accretion, while the PNS radius decreases due to the cooling by the neutrino (and also relativistic effect due to the mass increase). From this figure, one can observe that the PNS masses of 1D runs are heavier than the others and those with DD2 are heavier than those with TGTF. This ordering is correlated with the explodability of the models, i.e., a delayed shock revival results in more mass accretion to PNS. On the other hand, the radii of PNS in 1D models shrink faster than the others due to the heavier PNS masses they own. The radii with TGTF at the later epoch are smaller than those with DD2. This EOS dependence might come from the difference of the neutrino cooling that depends on the EOS stiffness and/or the detailed treatment of the nuclei. The average density also reflects all of these ingredients.

\begin{figure}[tbp]
\begin{center}
\includegraphics[scale=0.49]{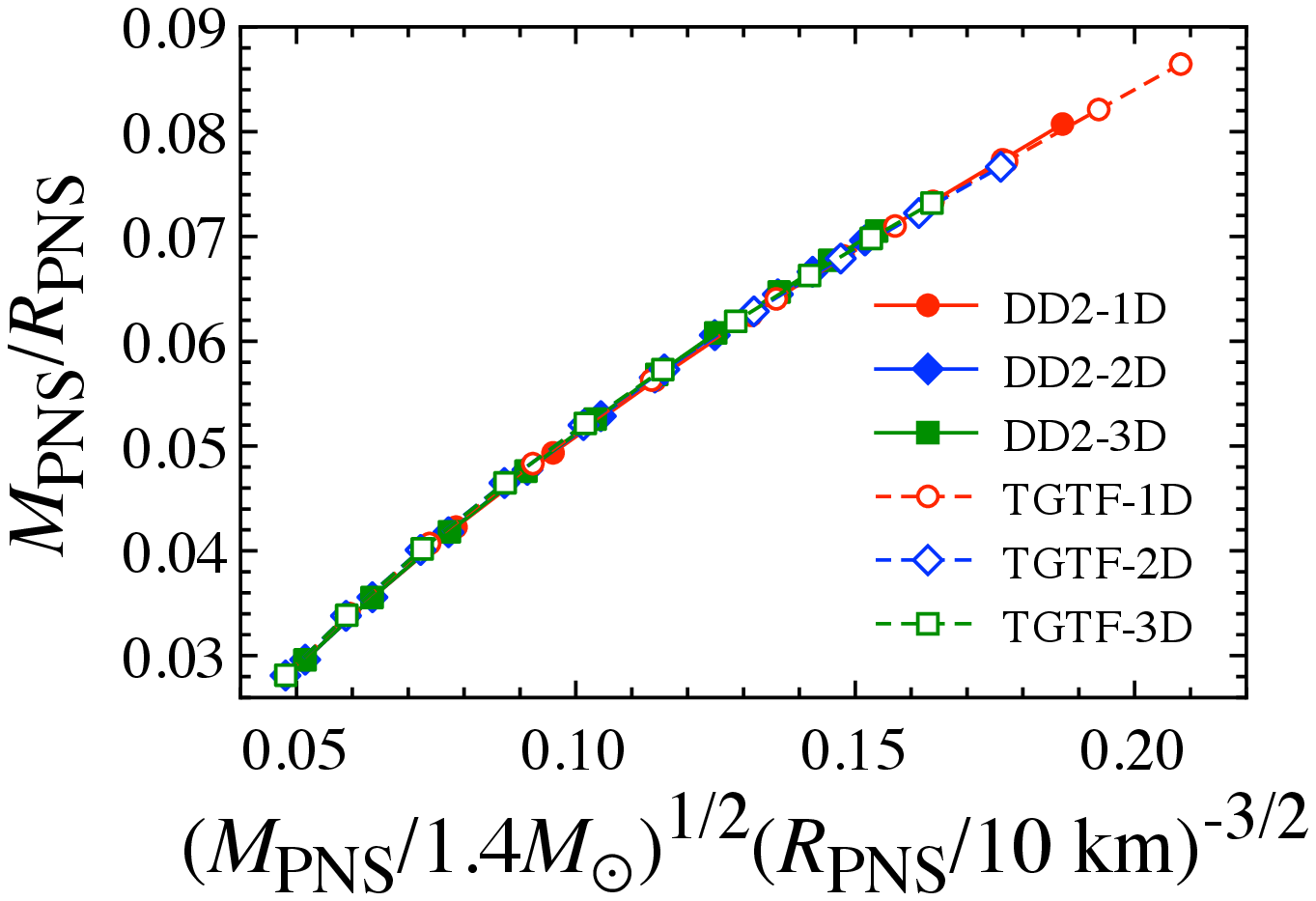}
\includegraphics[scale=0.49]{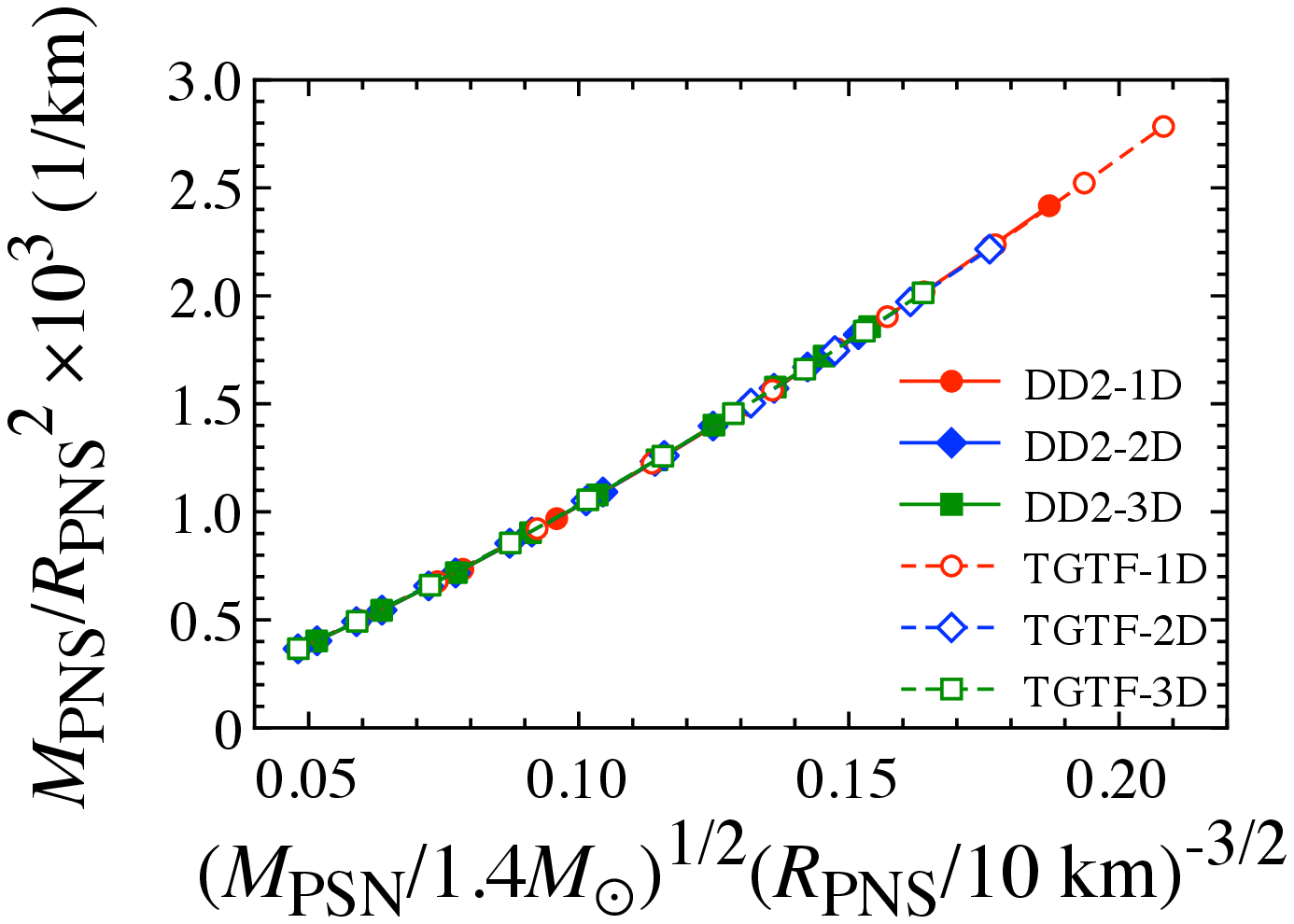}
\end{center}
\caption{
For several PNS models considered in this study, PNS compactness (left) and surface gravity (right) are shown as a function of the square root of normalized PNS average density. The filled marks with solid lines and open marks with dashed lines correspond to the PNS models constructed with DD2 and TGTF, respectively, while circles, diamonds, and squares correspond to the PNS models obtained from 1D, 2D, and 3D simulations. 
}
\label{fig:MR-ave}
\end{figure}

In the literature (e.g., Refs. \cite{SS2019,TCPOF2019b}), the PNS average density, compactness ($M_{\rm PNS}/R_{\rm PNS}$), and surface gravity ($M_{\rm PNS}/R_{\rm PNS}^2$) are considered to be able to characterize the oscillation modes. We surprisingly find that the relation between the PNS average density and compactness ($M_{\rm PNS}/R_{\rm PNS}$) are independent of not only the dimension in numerical simulations but also the adopted EOS at least during the period considered in this study. In the left panel of Fig. \ref{fig:MR-ave} we show the PNS compactness as a function of the square root of the normalized PNS average density. From this figure, one can observe that the PNS compactness almost becomes a linear function of the square root of the normalized PNS average density. It is expected that this degeneracy happens only in the early phase after the core bounce (or only for the PNS models with low compactness). Even so, this result may be a remarkable feature. We also find that, similar to the compactness, the surface gravity depends on only the average density, which is independent of the dimension of hydrodynamics and EOS, as shown in the right panel of Fig. \ref{fig:MR-ave}. We remark that these universal relations can be held even for the other PNS models at least for the region where the PNS compactness is less than $\sim 0.1$  (see Appendix \ref{sec:appendix_1}).
In general, these relations shown in Fig. \ref{fig:MR-ave} are not accepted for cold neutron stars. So, these relations must be a unique feature for PNSs, which may come from lower compactness of PNS, thermal effect in EOS for PNS, or other effects. Anyway, unfortunately we could not know the physical reason why these relations appear, but understanding of this feature would give us a new insight for PNSs.

We emphasize that one can use these relations for the parameter estimation. Previously we argued that simultaneous detection of two modes (e.g., $w$- and $f$- or $g$- mode) are necessary to extract $M_{\rm PNS}$ and $R_{\rm PNS}$ from the gravitational wave observation \cite{ST2016,SKTK2017}. However, owing to these relations shown in Fig. \ref{fig:MR-ave}, one can estimate $M_{\rm PNS}$ and $R_{\rm PNS}$ with the detection of only one mode in gravitational waves. 
That is, since $f$-mode frequency, which belongs to a kind of acoustic oscillations, can be expressed as a function of the PNS average density independently of the PNS EOS, as shown in Fig. \ref{fig:ff-ave}, one can know the value of $(M_{\rm PNS}/R_{\rm PNS}^3)^{1/2}$ via the observation of $f$-mode gravitational wave. Then, using this value, one can also know the value of $M_{\rm PNS}/R_{\rm PNS}$ (or $M_{\rm PNS}/R_{\rm PNS}^2$) via the relation shown in Fig. \ref{fig:MR-ave}. So, from the values of $(M_{\rm PNS}/R_{\rm PNS}^3)^{1/2}$ and $M_{\rm PNS}/R_{\rm PNS}$ (or $M_{\rm PNS}/R_{\rm PNS}^2$), one can determine the values of $M_{\rm PNS}$ and $R_{\rm PNS}$.
To derive more robust conclusion, we have to confirm the relation taking more samples.

\section{PNS oscillations}
\label{sec:Oscillation}

In order to determine the specific frequencies of PNS oscillations, we make a linear analysis on the PNS models given in the previous section. In this study, we simply adopt the relativistic Cowling approximation. That is, the metric perturbations are neglected during the fluid oscillations. Since we consider only the early phase after the core bounce in this study, where the PNS compactness is still relatively low, we believe that one can discuss the PNS oscillations even with the relativistic Cowling approximation. The perturbation equations and the boundary conditions are completely the same as in Ref. \cite{SKTK2019}, with which one can determine the eigenvalue ($\omega$) by solving the eigenvalue problem. Then, the frequency ($f$) is obtained via $f=\omega/(2\pi)$.   

There are several ways for the mode classification. In this study, we simply adopt the standard mode classification, which is the same as in the usual asteroseismology by counting the nodal number in the eigenfunctions. That is, the $f$-mode has no-node in the eigenfunction, while the $p_i$- and $g_i$- modes have $i$ nodes in their eigenfunctions. The frequencies of $p_i$-modes ($g_i$-modes) increase (decrease) as the nodal number $i$ increases, and the $f$-mode frequency is between the $g_1$- and $p_1$-mode frequencies. We remark that some of eigenmodes (especially with lower values of $i$) in the very early phase after core bounce may have the nodal number different from that in definition, because the eigenfunction may have more nodes in the vicinity of the stellar center. In that time, we identify such modes with using the shape of eigenfunctions, neglecting the additional nodal number in the vicinity of the stellar center.
We also remark that the results from most of multidimensional numerical simulations (e.g., \cite{Murphy09,MJM2013,CDAF2013,Yakunin15,KKT2016,Andresen16,OC2018,RMBVN19}) seem to show a strong $f$-mode (or the so-called surface $g$-mode) signal, but the $p$- and $g$- modes are less prominent. So, the extraction of the $p$- and $g$-modes from the observation data can be quite difficult with the current gravitational wave detectors. Even so, we believe that the theoretical prediction of $p$- and $g$-mode frequencies must be important, because they may be extracted due to the development of detector and/or of noise rejection in future.

\begin{figure*}[tbp]
\begin{center}
\includegraphics[scale=0.48]{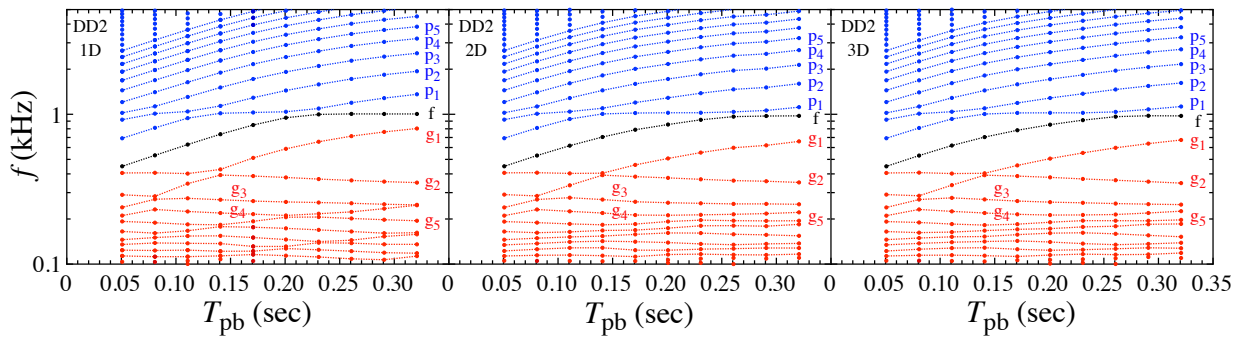} 
\end{center}
\caption{
Time evolution of eigenfrequencies in gravitational waves radiated from the PNS models of DD2-1D (left panel), DD2-2D (middle panel), and DD2-3D (right panel). In each panel, the $f$-, $g_i$-, and $p_i$-modes for $i=1-10$ are connected with dotted lines
}
\label{fig:ft-DD2123}
\end{figure*}

In Fig. \ref{fig:ft-DD2123}, we show the resultant frequency evolutions for the PNS models with DD2, where the left, middle, and right panels correspond to the results for 1D, 2D, and 3D simulations, respectively. As in Ref. \cite{SS2019}, in every case one can clearly observe the phenomena of the avoided crossing. We remark that we could not find any $g$-mode oscillations in our previous study \cite{SKTK2019}, while this time we can find the $g$-mode oscillations even though we adopt the same numerical code for determining the specific frequencies of the PNS models. This may come from the fact that our previous background model shows strong feature of Standing Accretion Shock Instability (SASI), while our current model does not show a prominent signature of SASI activity. That is, 
due to the existence of strong SASI, some periodic push induces the mixing of $Y_e$ and entropy inside the PNS and can flatten the stratification of them significantly. The modified distribution of  $Y_e$ and entropy may suppress the $g$-mode oscillations, because the buoyancy of the stratified media is restoring force of the $g$-modes. This may be a reason why the $g$-mode could not be found in the previous study.
Anyway, from Fig. \ref{fig:ft-DD2123} we can find that the frequency evolution obtained from the PNS model with 1D simulation (left panel) is different from that with multi-dimensional simulations (middle and right panels).

\begin{figure*}[tbp]
\begin{center}
\begin{tabular}{cc}
\includegraphics[scale=0.5]{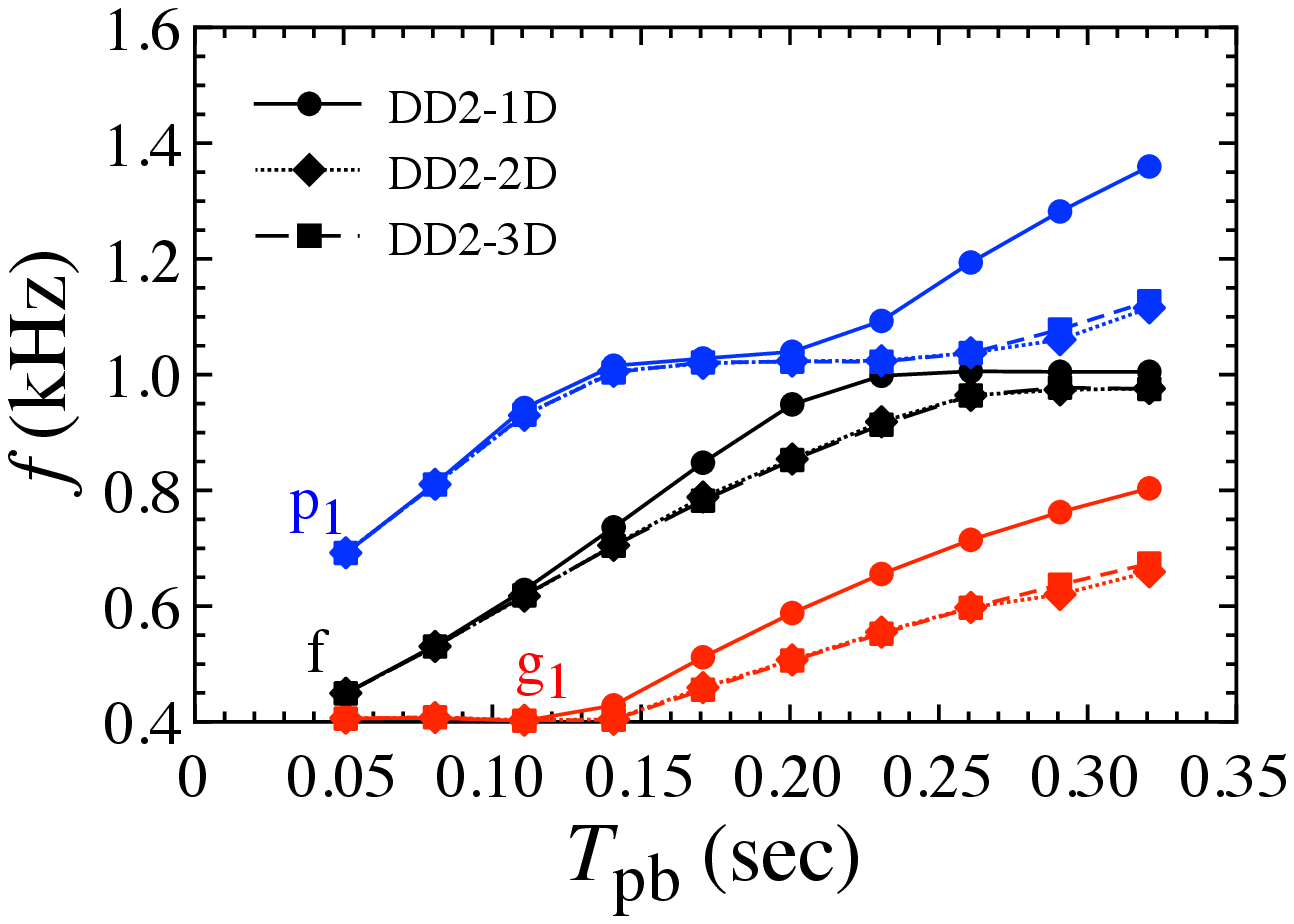} &
\includegraphics[scale=0.5]{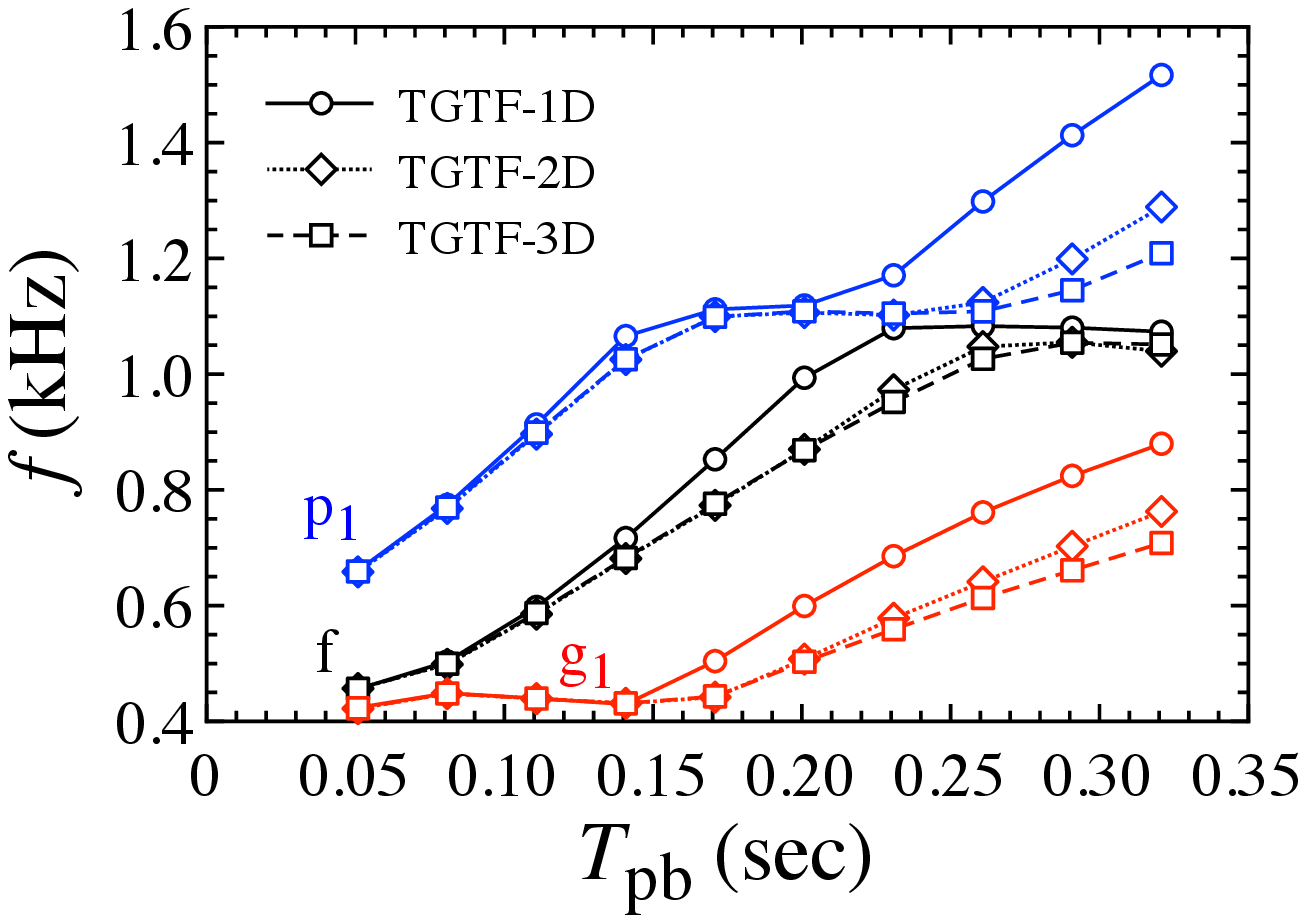} 
\end{tabular}
\end{center}
\caption{
Time evolution of the $p_1$-, $f$-, and $g_1$-mode gravitational waves from the PNS models constructed with DD2 in the left panel and TGTF in the right panel. In each panel, circles, diamonds, and squares correspond to the results for the PNS models obtained from 1D, 2D, and 3D simulations.
}
\label{fig:ft}
\end{figure*}

To clarify this point, we show the frequency evolutions of the $f$-, $p_1$-, and $g_1$-mode gravitational waves are shown in Fig. \ref{fig:ft}, where circles, diamonds, and squares correspond to the results obtained by the PNS models with 1D, 2D, and 3D simulations, respectively, while the left and right panels correspond to the results for the PNS models with DD2 and TGTF. From this figure, one can observe that the time when the avoided crossing happens between the $f$- and $p_1$-modes for the PNS model with 1D simulation is completely different from that for the PNS model with multi-dimensional simulations. Meanwhile, we also find that the results for the PNS model with 2D simulation are very similar to those with 3D simulations. In addition, as mentioned in Ref. \cite{SS2019}, one also observes the plateau in the $f$-mode frequency due to the avoided crossing to the $p_1$-mode, even though the PNS mass and radius are still changing as in Fig. \ref{fig:MRave}. It is expected that this plateau would continue up to the time when the avoided crossing to the $g_1$-mode would happen \cite{SS2019}.

\begin{figure}[tbp]
\begin{center}
\includegraphics[scale=0.5]{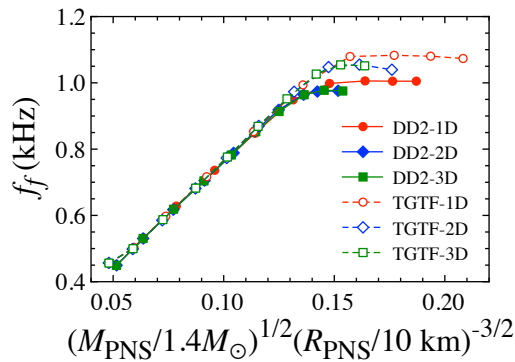} 
\end{center}
\caption{
Frequency of $f$-mode gravitational waves is shown as a function of the square root of normalized PNS average density. The filled and open marks correspond to the results for the PNS models with DD2 and TFTG, respectively, while the circles, diamonds, and squares denote the results for the PNS models obtained via the 1D, 2D, and 3D simulations.  
}
\label{fig:ff-ave}
\end{figure}

On the other hand, as shown in Fig. \ref{fig:ff-ave} we find that the relation between the $f$-mode frequency and the square root of the normalized PNS average density is basically independent of the dimension of numerical simulations for preparing the PNS models. That is, as a function of the square root of the normalized PNS average density, one can discuss the $f$-mode frequency in the PNSs obtained via multi-dimensional simulations even with that in the PNSs with 1D simulations. Additionally, as in the previous studies (e.g., \cite{ST2016,SKTK2017,SKTK2019,SS2019}), one can observe from Fig. \ref{fig:ff-ave} that the $f$-mode frequency are almost independent of the EOS as a function of the square root of the average density, although one can also see the deviation in the $f$-mode frequency plateau. This deviation seems to be more or less consistent with the results in Ref. \cite{SS2019}.

\begin{figure*}[tbp]
\begin{center}
\begin{tabular}{cc}
\includegraphics[scale=0.5]{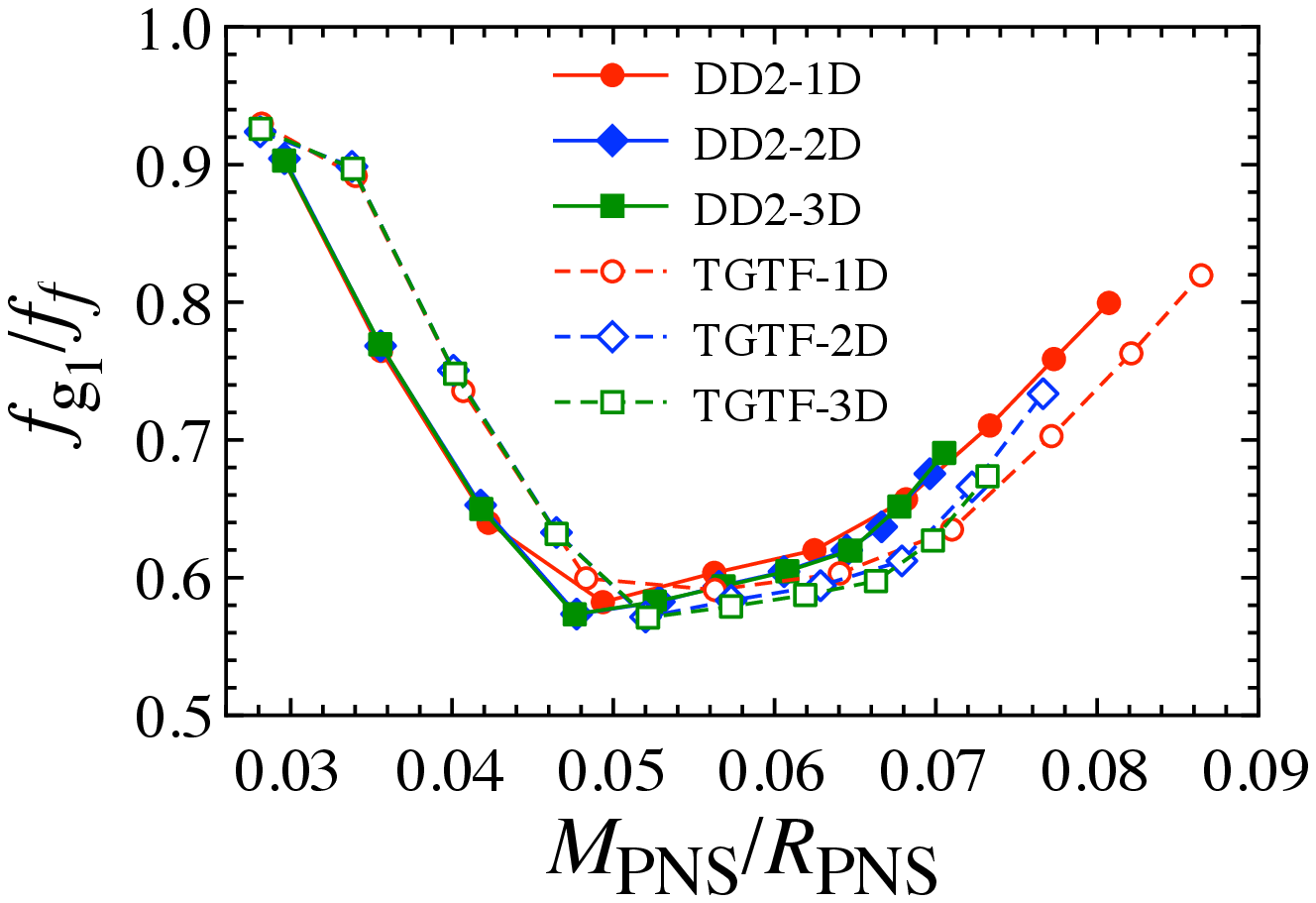} & 
\includegraphics[scale=0.5]{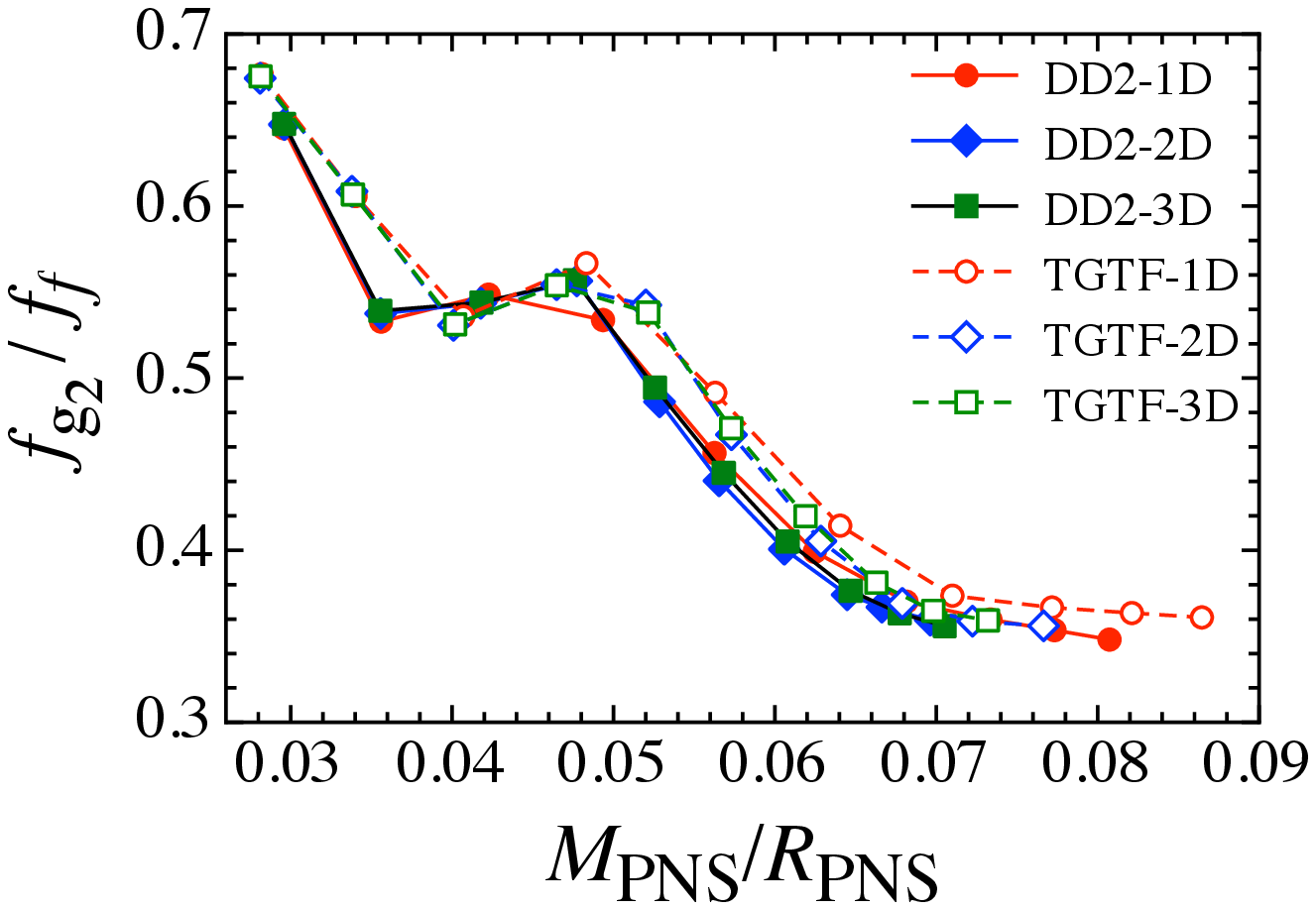} 
\end{tabular}
\end{center}
\caption{
Ratio of the $g_1$-mode frequency to the $f$-mode frequency is shown as a function of PNS compactness in the left panel, while the ratio of the $g_2$-mode frequency to the $f$-mode frequency is shown in the right panel.
}
\label{fig:gf-MR}
\end{figure*}

\begin{figure}[tbp]
\begin{center}
\begin{tabular}{cc}
\includegraphics[scale=0.5]{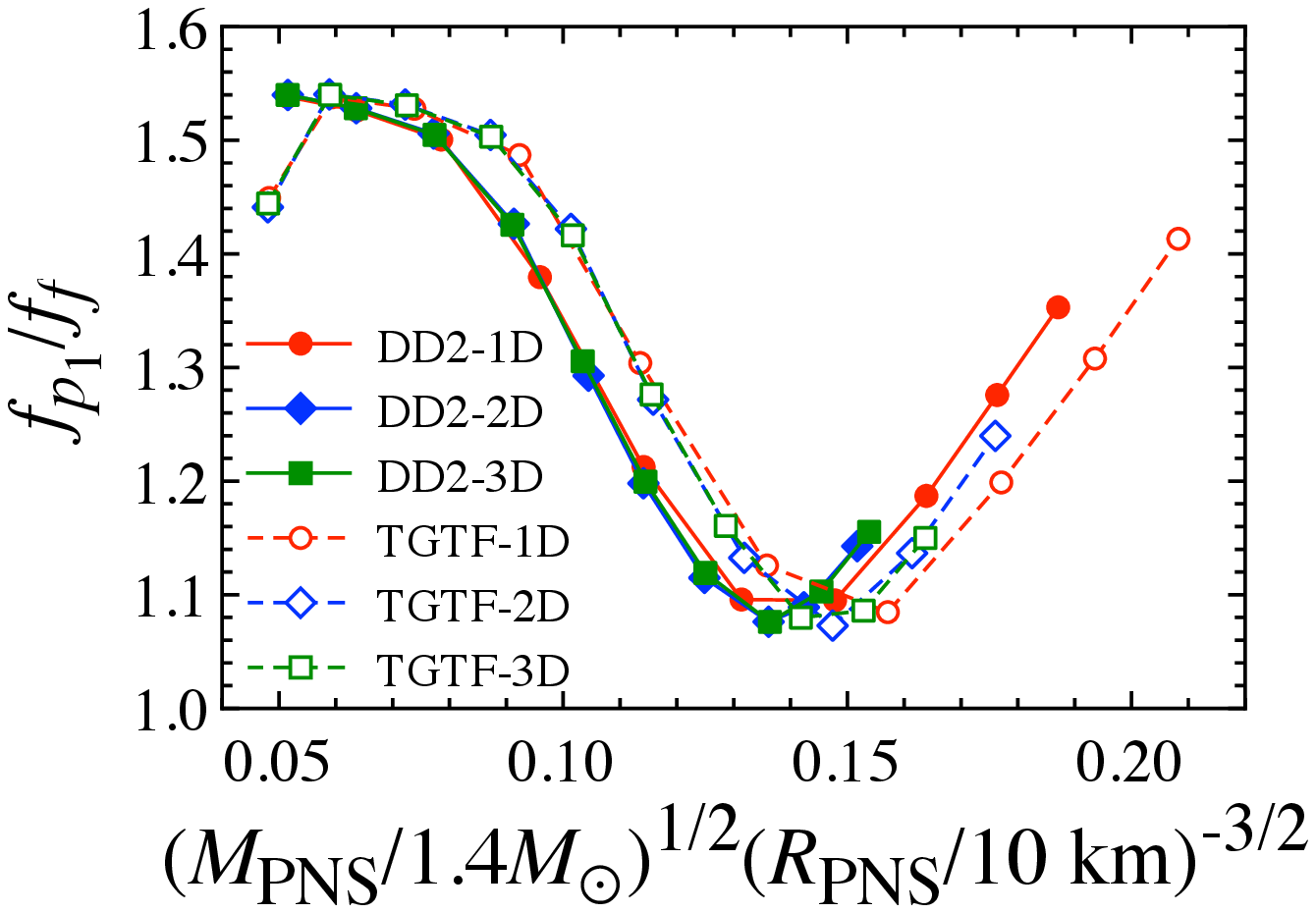} & 
\includegraphics[scale=0.5]{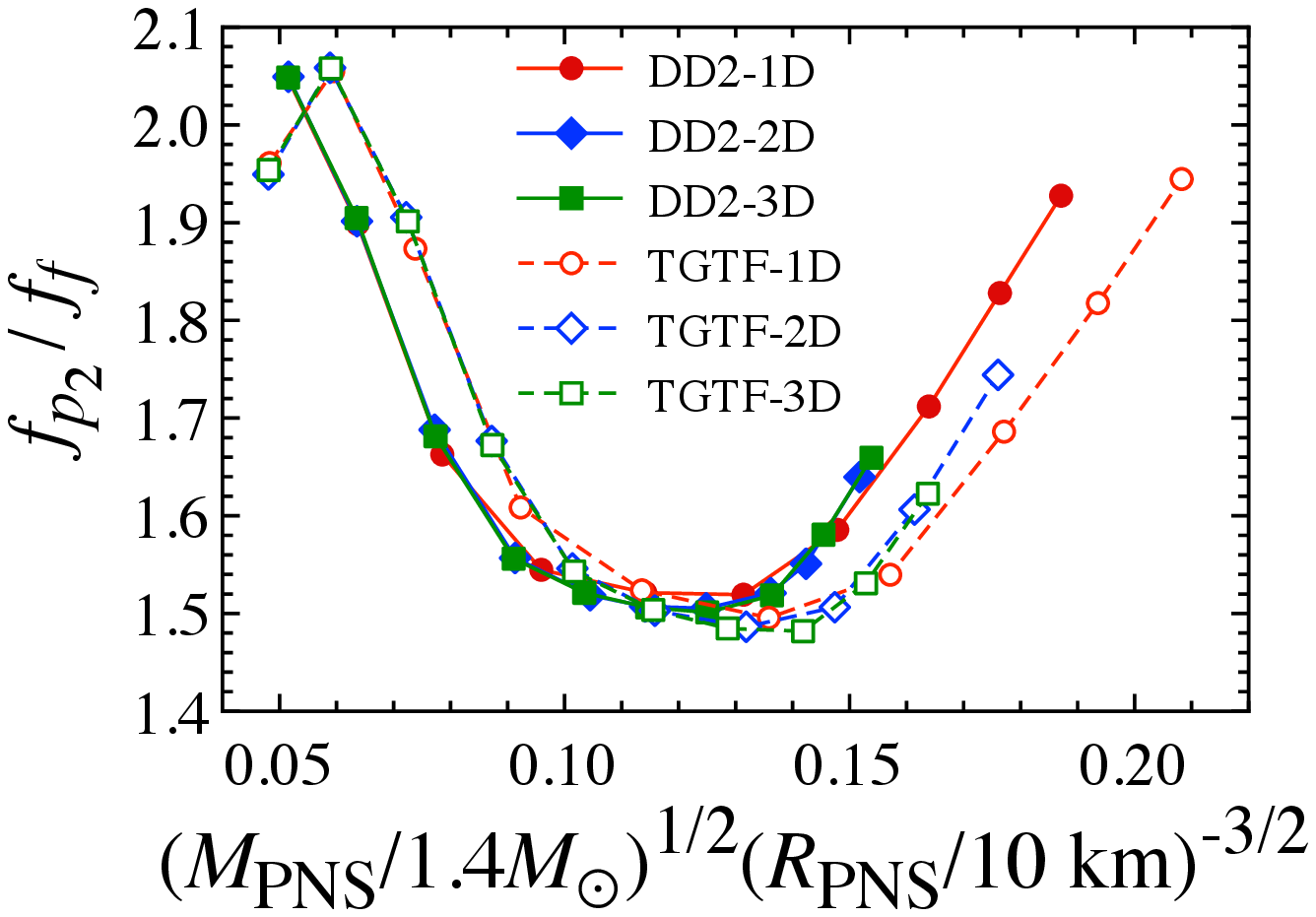} 
\end{tabular}
\end{center}
\caption{
Ratio of the $p_1$-mode frequency to the $f$-mode frequency is shown as a function of square root of normalized PNS average density in the left panel, while the ratio of the $p_2$-mode frequency to the $f$-mode frequency is shown in the right panel.
}
\label{fig:pf-ave}
\end{figure}

Furthermore, as shown in the left panel of Fig. \ref{fig:gf-MR}, we can confirm that the ratio of the $g_1$-mode frequency to the $f$-mode frequency, $f_{g_1}/f_f$, is strongly associated with the PNS compactness almost independently of the PNS EOS, which is originally pointed out in Ref. \cite{SS2019}. In this study, we newly find that the ratio of the $g_2$-mode frequency to the $f$-mode frequency, $f_{g_2}/f_f$, also strongly depends on the PNS compactness, where the dependence on the EOS is very weak (see the right panel of Fig. \ref{fig:gf-MR}). In addition, we find that such ratios are almost independent of the dimension of the numerical simulation for preparing the PNS models, once one selects an EOS. Similarly, as shown in the left panel of Fig. \ref{fig:pf-ave} we find that the ratio of the $p_1$-mode frequency to the $f$-mode frequency, $f_{p_1}/f_f$, is strongly correlated to the PNS average density. The dependence of this frequency ratio on the PNS properties has been originally pointed out in Ref. \cite{Camelio17}, where the correlation with the quantity defined by
\begin{equation}
  Q_0 \equiv \frac{1}{f_f}\left(\frac{M_{\rm PNS}}{M_\odot}\right)^{1/2}\left(\frac{R_{\rm PNS}}{10\ {\rm km}}\right)^{-3/2} \label{eq:Q0}
\end{equation}
is shown independently of the progenitor mass and EOS. In this study, we show the correlation with the square root of the PNS average density instead of $Q_0$. In fact, one can obviously see the strong correlation between $f_{p_1}/f_f$ and PNS average density, which weakly depends on the EOS, even thought the behavior of $f_{p_1}/f_f$ is not monotonically. In the similar way, we also find the strong correlation between $f_{p_2}/f_f$ and PNS average density, as shown in the right panel of Fig. \ref{fig:pf-ave}. In any case, one can observe that the both of $f_{p_1}/f_f$ and $f_{p_2}/f_f$ are almost independent of the dimension of numerical simulations for preparing the PNS models, adopting a specific EOS. Finally, we have to mention that one can plot the similar figures to Figs. \ref{fig:ff-ave} -- \ref{fig:pf-ave} as a function of either the PNS compactness or the square root of the PNS average density at least during the time interval we considered in this study, because the PNS compactness almost behaves as a linear function of the square root of the PNS average density independently of the EOS and dimension of the numerical simulations, as shown in Fig. \ref{fig:MR-ave}. Even so, Figs. \ref{fig:ff-ave} and \ref{fig:gf-MR} are plotted as in the previous studies, while Fig. \ref{fig:pf-ave} is plotted as a function of the square root of the PNS average density because $p_i$-modes are a kind of acoustic waves.

\section{Conclusion and discussion}
\label{sec:Conclusion}

In this article, we examined the eigenfrequencies of the gravitational waves from the PNSs in the early phase after the core bounce, especially focusing on how the frequencies depend on the dimension of numerical simulations for preparing the PNS models. For this purpose, we performed 1D, 2D, and 3D numerical simulations for core-collapse supernovae with the same numerical scheme, where we adopted the $20M_\odot$ progenitor model and two EOSs. As a result, we showed that the time evolution of the gravitational wave frequencies depends on the dimension of numerical simulations and the EOS, while we also found that the frequency evolutions for the PNS models with 2D simulations are very similar to those with 3D simulations. Even so, focusing on the relation between the $f$-mode frequency and the PNS average density, the relation between the ratio of the $g_1$-mode ($g_2$-mode) frequency to the $f$-mode frequency and the PNS compactness, and the relation between the ratio of the $p_1$-mode ($p_2$-mode) frequency to the $f$-mode frequency and the PNS average density, we found that these are independent of the dimension of the numerical simulations. Thus, even with the frequency calculated for the PNS models with 1D simulations, one can safely discuss such dependence in the gravitational waves. In addition, we confirmed and newly found that these dependence of the $f$-mode frequency and the combinations of frequencies on the PNS properties are almost independent of the PNS EOS. Furthermore, we accidentally found that the relation between the PNS compactness (or the PNS surface gravity) and PNS average density for the early phase after the core bounce are independent of not only the dimension of the numerical simulation for preparing the PNS models but also the PNS EOS. Owing to these relations, one could extract the PNS mass and radius at each time after core bounce via the observation of only the $f$-mode gravitational waves.

In this study, we focus only on asteroseismology with the PNSs provided by successful core-collapse supernovae, but one can similarly make an analysis with the PNSs for failed supernovae, where a  massive progenitor would directly collapse to a black hole without explosion \cite{SS2019}. For the case of failed supernovae, the mass of PNS reaches the maximum mass and the neutrino signal disappears at the moment when the PNS collapse to a black hole. So, one could discuss the PNS with the maximum mass through the gravitational wave observations with the help of the neutrino observations. Unlike the case of successful core-collapse supernovae, for failed supernovae one would detect gravitational waves and neutrino without any electromagnetic signals. This type of observation would become a signature for failed supernovae.

\acknowledgments

This work is supported in part by Grant-in-Aid for Scientific Research of Japan Society for the Promotion of Science (JSPS, Nos. 
JP17K05458, 
JP17H01130, 
JP17K14306, 
JP18H01212, 
JP19KK0354 
),
the Ministry of Education, Science and Culture of Japan (MEXT, Nos.
JP17H06357, 
JP17H06364, 
JP17H05206, 
JP20H04753 
),
Joint Institute for Computational Fundamental Science (JICFuS) as a priority issue to be tackled by using Post `K' Computer.
This work is also supported by the National Institutes of Natural Sciences 
(NINS) program for cross-disciplinary
study (Grant Numbers 01321802 and 01311904) on Turbulence, Transport,
and Heating Dynamics in Laboratory and Solar/Astrophysical Plasmas:
"SoLaBo-X?.
Numerical computations were in part carried out on Cray XC50, PC cluster and analysis server at Center for Computational Astrophysics, National Astronomical Observatory of Japan.

\appendix
\section{Comparison with the other PNS models}   
\label{sec:appendix_1}

In this study, we show a kind of universality in the relation between PNS compactness and PNS average density in the left panel of Fig. \ref{fig:MR-ave} and the relation between the PNS surface gravity and PNS average density in the right panel of Fig. \ref{fig:MR-ave}. Unfortunately, so far we could not find the physical reason why this type of universal relation is established. So, in order to corroborate our finding, we reconsider the same relation by adding the other PNS models, instead of the physical reason. For this purpose, we consider the PNS models discussed in Ref. \cite{SS2019}. In fact, the PNSs discussed in Ref. \cite{SS2019} eventually collapse to black holes, when the PNS mass is over the maximum mass allowed from the adopted EOS. Thus, since the PNS models in Ref. \cite{SS2019} are more massive (or more compact) than those discussed in this study, one can expect to see the behavior of the relations shown in Fig. \ref{fig:MR-ave} with wider range.

\begin{figure}[tbp]
\begin{center}
\includegraphics[scale=0.49]{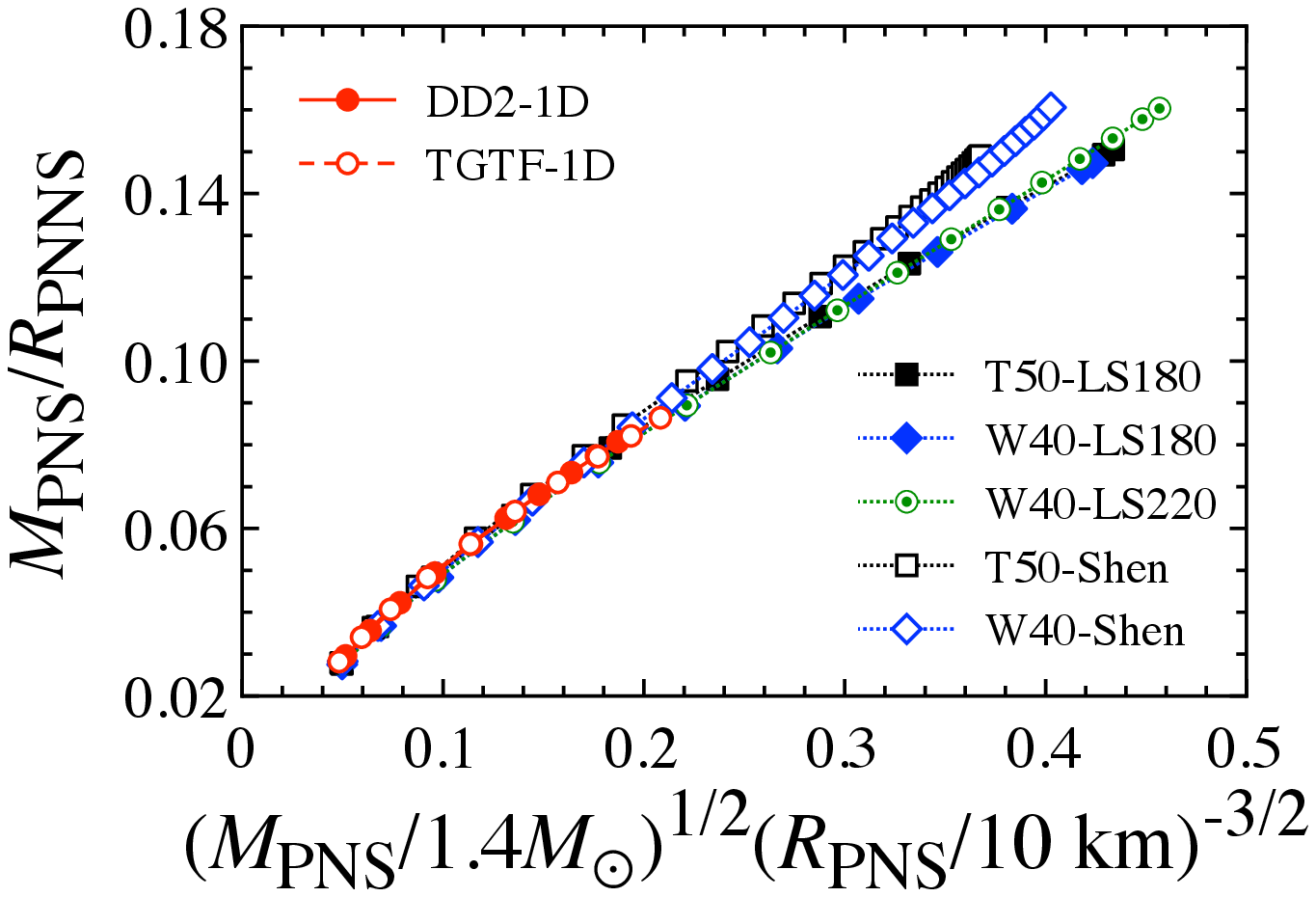}
\includegraphics[scale=0.49]{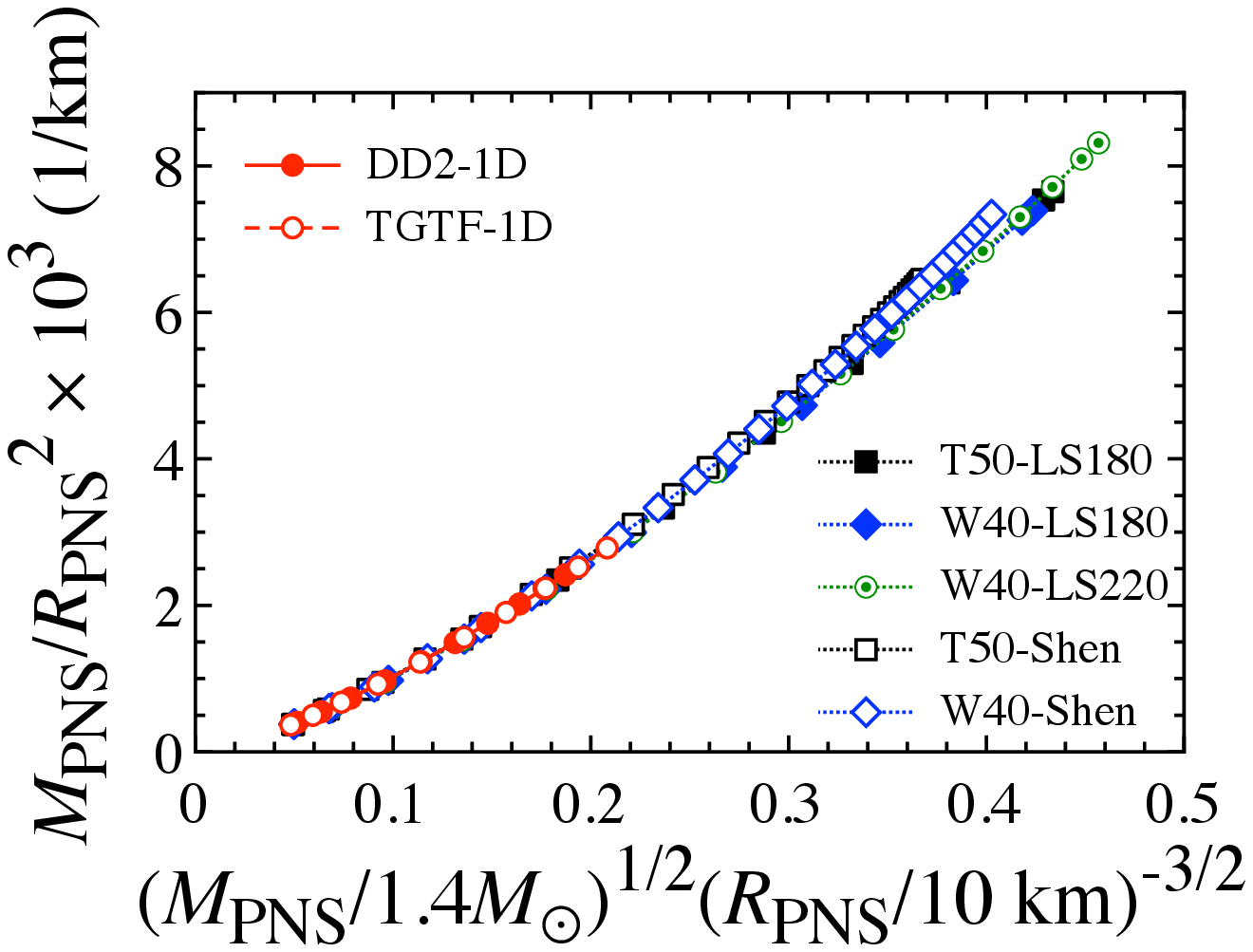}
\end{center}
\caption{
The PNS compactness and the PNS surface gravity are shown as a function of the square root of the normalized PNS average density in the left and right panel, respectively. 
}
\label{fig:MR-ave-SS2019}
\end{figure}

The PNS models in Ref. \cite{SS2019} are produced with the data obtained by the 1D general relativistic neutrino-radiation hydrodynamics simulation, where the neutrino transfer is solved by the Boltzmann equation in general relativity \cite{Yamada97,Yamada99,Sumiyoshi05}. In the simulatios, the progenitor models are adopted a $40M_\odot$ star based on Ref. \cite{WW95} and a $50M_\odot$ star based on Ref. \cite{TUN07} (refereed to as W40 and T50 hereafter), while three different EOSs are considered, i.e., the Shen EOS \cite{Shen_EOS}, LS180, and LS220 \cite{LS_EOS}. Then, the resultant PNS models (T50-LS180, W40-LS180, W40-LS220, T50-Shen, and W40-Shen) are shown in Fig. \ref{fig:MR-ave-SS2019} together with the PNS models discussed in this study, i.e., DD2-1D and TGTF-1D. 

From this figure, the universal relation between the PNS compactness and the PNS average density can be seen at least for $M_{\rm PNS}/R_{\rm PNS} \lsim 0.1$. On the other hand, the universal relation between the PNS surface graivity and the PNS average density can be seen in wider range. 

\end{document}